\documentclass[aps,prc,twocolumn,showpacs,showkeys,superscriptaddress,nofootinbib]{revtex4-2}
\usepackage{graphicx}
\usepackage{amsmath,amssymb}
\usepackage{bm}
\usepackage{color}
\usepackage{multirow}
\usepackage{mathrsfs}
\usepackage{changes}
\usepackage{hyperref}
\hypersetup{
    colorlinks = true,
    linkcolor = blue,
    citecolor = blue,
    urlcolor = blue
}
\date{\today}

\begin{document}
\title{Low-energy $^{3}\mathrm{He}(\alpha,\gamma)^{7}\mathrm{Be}$ reaction within the Skyrme potential framework}

    \author{Nguyen Le Anh}
	\email{anhnl@hcmue.edu.vn}
	\affiliation{Department of Physics, Ho Chi Minh City University of Education, 280 An Duong Vuong, Cho Quan Ward, Ho Chi Minh City, Vietnam}

    \author{Nguyen Gia Huy}
    \affiliation{Department of Physics, Ho Chi Minh City University of Education, 280 An Duong Vuong, Cho Quan Ward, Ho Chi Minh City, Vietnam}

    \author{Dao Nhut Anh}
    \affiliation{Department of Physics, Ho Chi Minh City University of Education, 280 An Duong Vuong, Cho Quan Ward, Ho Chi Minh City, Vietnam}
    
    \author{Do Huy Tho}
    \affiliation{Department of Physics, Ho Chi Minh City University of Education, 280 An Duong Vuong, Cho Quan Ward, Ho Chi Minh City, Vietnam}
    
    \author{Hoang Thai An}
    \affiliation{Department of Physics, Ho Chi Minh City University of Education, 280 An Duong Vuong, Cho Quan Ward, Ho Chi Minh City, Vietnam}

\begin{abstract}
\noindent
\textbf{Background:} The $^{3}\mathrm{He}(\alpha,\gamma)^{7}\mathrm{Be}$ reaction plays a crucial role in the proton-proton chain and Big Bang nucleosynthesis, affecting solar neutrino fluxes and primordial element abundances. Experimental data at astrophysical energies remain uncertain due to the extremely low cross sections. \\
\textbf{Purpose:} This work uses a microscopic potential-model approach to construct the $^{3}\mathrm{He}+\alpha$ potential from the $\text{nucleon}+\alpha$ interaction, aiming to describe low-energy elastic scattering and to calculate the astrophysical $S$ factor of the $^{3}\mathrm{He}(\alpha,\gamma)^{7}\mathrm{Be}$ reaction. \\
\textbf{Method:} The nucleon-nucleus potential is derived from self-consistent Skyrme Hartree-Fock (HF) calculations extended to the continuum. The $^{3}\mathrm{He}+\alpha$ potential is then obtained by folding the HF potential with the $^{3}\mathrm{He}$ density. A small number of scaling parameters is constrained by elastic-scattering data.\\
\textbf{Result:} The scaled Skyrme HF potential and folded potential simultaneously reproduce the low-energy $p+\alpha$ and $^{3}\mathrm{He}+\alpha$ $s$-wave phase shifts, respectively. The calculated astrophysical $S$ factor of $^{3}\mathrm{He}(\alpha,\gamma)^{7}\mathrm{Be}$ shows good agreement with experimental data, yielding the recommended value $S_{34}(0) = 0.610 \pm 0.024$~keV~b. A moderate sensitivity of $S_{34}(0)$ to the choice of projectile density is also observed in the folding procedure. \\
\textbf{Conclusion:} The Skyrme HF-based potential provides a unified and predictive microscopic framework for describing both elastic scattering and radiative capture in light nuclei.
\end{abstract}

\maketitle

\section{Introduction} \label{sec:introduction}

Over the past several decades, solar physics has faced persistent discrepancies between theoretical solar models and observational constraints \cite{adelberger1998,adelberger2011,acharya2025}. In particular, the predicted solar neutrino fluxes deviate slightly from measured values, with the differences being susceptible to the reaction rate of the $^{3}\mathrm{He}(\alpha,\gamma)^{7}\mathrm{Be}$ reaction, a key process in the proton-proton chain of stellar hydrogen burning \cite{cyburt2008}. Accurately determining the reaction rate, therefore, remains a central objective of both experimental measurements and theoretical models in nuclear astrophysics.

Experimental studies of this reaction are challenging due to the strong Coulomb barrier, which leads to extremely low cross sections and large uncertainties at astrophysical energies. Various measurements have been reported in Refs.~\cite{costantini2007,costantini2008,osborne1984,brown2007,nagatani1969,dileva2009,hilgemeier1988,parker1963,bordeanu2013}, but the results remain somewhat inconsistent. The energy ranges relevant to solar models and Big Bang nucleosynthesis (BBN) are below 500~keV in the center-of-mass (c.m.) frame \cite{takacs2015}. Measurements in these regions, particularly those in Refs.~\cite{costantini2007,costantini2008,brown2007,osborne1984,nagatani1969}, are therefore crucial for benchmarking theoretical models of low-energy nuclear reactions.

A wide range of theoretical approaches has been developed to evaluate the astrophysical $S$ factor of this reaction, including the $R$-matrix method~\cite{kontos2013,deboer2014}, cluster models~\cite{kajino1986,buck1985,solovyev2019}, shell-model calculations~\cite{dohet2011,dohet2016,vorabbi2019,atkinson2025}, fermionic molecular dynamics model~\cite{neff2011}, effective field theory~\cite{higa2018,premarathna2020,zhang2020}, and potential models~\cite{mohr1993,mohr2009,xu2013,dubovichenko2017,tursunov2018}. Each approach has its own merits; however, the potential model remains a robust framework for reproducing experimental data. An important advantage is that low-energy direct capture is dominated by long-range electric dipole ($E1$) transitions, allowing the reaction to be described accurately once the scattering phase shifts and bound-state asymptotics are constrained~\cite{tursunov2018}. In particular, the microscopic potential model, which derives the interaction from effective nucleon-nucleon forces, provides a self-consistent foundation with improved predictive power compared to purely phenomenological approaches in Refs.~\cite{xu2013,dubovichenko2017,tursunov2018}.

The microscopic potential model starts from an effective nucleon-nucleon interaction and applies suitable approximations to treat the nuclear many-body problem. Although certain adjustments are often necessary, the model aims to minimize the number of free parameters and retain a strong connection to the underlying nuclear interaction. In the present work, the potential is derived self-consistently from the Skyrme Hartree-Fock (HF) framework, providing a unified description of both bound and scattering states \cite{dover1971,dover1972}. This approach, previously applied to nucleon-nucleus elastic scattering \cite{anh2022Be10,anh2023O1415} and nucleon radiative capture \cite{anh2021CNO,anh2021C12O16,anh2022Li7Be7}, is extended here for the first time to describe low-energy nucleus-nucleus systems.

The HF calculation is performed using the SLy4 interaction to obtain the single-particle states and the self-consistent mean-field potential after convergence \cite{vautherin1972,colo2013}. The resulting HF potential is extended to the continuum region to describe scattering states corresponding to unoccupied states. The Skyrme interaction parameters are kept unchanged, and only the converged nuclear potentials are scaled. A scaling factor $\lambda_0$ is introduced to slightly adjust the central potential to reproduce the experimental phase shifts of $p+\alpha$ elastic scattering. The nucleon-$\alpha$ potential is then folded with the density distribution of $^{3}\mathrm{He}$ to construct the $^{3}\mathrm{He}+\alpha$ potential, which is used to analyze the corresponding phase shifts and elastic scattering cross sections. Comparison of the $\lambda_0$ values extracted from both systems serves as a test of the consistency of the model within the same microscopic framework.

The HF potential in combination with a folding approach is employed to calculate the $E1$ transition in the $^{3}\mathrm{He}(\alpha,\gamma)^{7}\mathrm{Be}$ radiative-capture reaction, where the $E1$ transition dominates at astrophysical energies ~\cite{mohr1993,tursunov2018}. At such low energies, the primary contribution arises from $s$-wave scattering to bound $p$ states, making the scaling factor $\lambda_0$ a crucial input for accurately describing the capture process.

The paper is organized as follows. Section~\ref{sec:method} presents the theoretical framework of the microscopic potential model, including the formalism for radiative-capture cross sections and the construction of the Skyrme-based potential. Section~\ref{sec:results} discusses the results for low-energy $p+\alpha$ and $^{3}\mathrm{He}+\alpha$ elastic scattering and the astrophysical $S$ factor of the $^{3}\mathrm{He}(\alpha,\gamma)^{7}\mathrm{Be}$ reaction. The main conclusions are summarized in Section~\ref{sec:conclusion}.

\section{Method of calculation} \label{sec:method}
\subsection{Radiative capture cross section in potential model approach} \label{subsec:XS}

In low-energy charged-particle reactions, the cross section $\sigma(E)$ decreases steeply with decreasing energy due to the suppression of the Coulomb barrier penetration. To remove this strong energy dependence, the astrophysical $S$ factor is introduced as
\begin{equation} \label{eq:S-factor}
    S(E) = E \sigma(E) \exp(2\pi \eta),
\end{equation}
where $\eta$ is the Sommerfeld parameter that depends on the charges, reduced mass, and relative energy of the interacting nuclei. In stellar nucleosynthesis, $S(E)$ is typically labeled by the mass numbers of the interacting nuclei; for instance, $S_{34}(E)$ corresponds to the $^{3}\mathrm{He}(\alpha,\gamma)^{7}\mathrm{Be}$ reaction.

For a direct radiative-capture process $a + A \to B + \gamma$ dominated by the electric dipole ($E1$) transition, the cross section can be written as
\begin{equation} \label{eq:sigma}
\sigma(E) = \dfrac{16\pi}{9} \dfrac{k_\gamma^3}{\hbar v} \dfrac{1}{(2J_a+1)(2J_A+1)} \left| M_{E1} \right|^2,
\end{equation}
where $v$ is the relative velocity between the projectile ($a$) and the target ($A$), $k_\gamma$ is the wave number of the emitted photon, and $J_a$ and $J_A$ are their spins.

The $E1$ transition matrix element $M_{E1}$ can be factorized into a single-particle matrix element and angular momentum coupling terms as
\begin{equation} \label{eq:M_E1}
|M_{E1}| = e_{\text{eff}}\, \hat{J}_S\, \hat{J}_B \left\{ \begin{matrix}
    j_b & J_B & J_A \\
    J_S & j_s & 1
\end{matrix} \right\} |M^{\text{s.p.}}_{E1}|,
\end{equation}
where $\hat{J} = \sqrt{2J + 1}$ is the standard angular-momentum factor, and the curly bracket denotes the Wigner $6j$ symbol. The effective charge is obtained from the cluster model expression $e_{\text{eff}} = e (Z_a/m_a - Z_A/m_A) \mu$, where $\mu$ is the reduced mass of the projectile-target system~\cite{tursunov2018}. The effective charge is $e_{\text{eff}} = 0.281e$ for the $^{3}\mathrm{He}(\alpha,\gamma)^{7}\mathrm{Be}$ reaction. 
$J_S$ and $J_B$ are the total angular momenta of the projectile-target system in the scattering and bound channels, respectively, with $J_B$ corresponding to the spins of $^{7}\mathrm{Be}$ in the bound state. The total angular momenta are defined as $\vec{j}_s = \vec{\ell}_s + \vec{J}_a$ and $\vec{j}_b = \vec{\ell}_b + \vec{J}_a$, where $\ell_s$ and $\ell_b$ are the relative orbital angular momenta in the scattering and bound states, respectively. Finally, $\vec{J}_S = \vec{j}_s + \vec{J}_A$ and $\vec{J}_B = \vec{j}_b + \vec{J}_A$ define the total angular momenta of the compound system in each channel.

The single-particle transition matrix element is written as
\begin{equation} \label{eq:Msp_E1}
|M^{\text{s.p.}}_{E1}| = \sqrt{\dfrac{3}{4\pi}}\, \hat{\ell}_s\, \hat{j}_s\, \hat{j}_b\, (\ell_s\, 0\, 1\, 0 \mid \ell_b\, 0) \left\{ \begin{matrix}
    \ell_s & J_a & j_s \\
    j_b & 1 & \ell_b
\end{matrix} \right\} I,
\end{equation}
where the round bracket denotes the Clebsch-Gordan coefficient, and the radial overlap integral $I$ is defined as
\begin{equation} \label{eq:I}
I = \int u_{\ell_b j_b}(r)\, u_{\ell_s j_s}(E, r)\, r\, dr,
\end{equation}
where $u_{\ell_b j_b}(r)$ and $u_{\ell_s j_s}(E, r)$ are the bound and scattering radial wave functions in the c.m. frame. These wave functions are obtained within a two-body cluster model by solving the radial Schr\"odinger equation for the $^{3}\mathrm{He}+\alpha$ system using the folded HF potentials. According to Ref.~\cite{tursunov2018}, convergent results were reported for $r_{\max}=40$~fm at low astrophysical energies. The equation is therefore solved up to $r_{\max}=50$~fm with a step size of $0.05$~fm. Scattering solutions are matched to Coulomb functions, and bound-state solutions are normalized to Whittaker asymptotics.

\subsection{Microscopic potential derived from the Skyrme Hartree-Fock approach} \label{subsec:OP}
The nucleus-nucleus potential can be obtained using the single-folding procedure applied to the nucleon-nucleus potential \cite{kuprikov2016}
\begin{equation} \label{eq:folding}
    V_{aA}(E,r) = \sum_{q=n,p}\int  \rho^{(a)}_q(\vec{r}_a) V_q(E,s) \,d^3 r_a,
\end{equation}
where $\vec{r}_a$ is the position of a nucleon inside the projectile measured from its center of mass, $\vec{R}$ connects the centers of the projectile and target nuclei, and $\vec{s} = \vec{r}_a - \vec{R}$ denotes the distance between a nucleon in the projectile and the target center.

The projectile density $\rho_q^{(a)}(r)$ is described by a two-parameter Fermi (2pF) distribution \cite{myers1973,aygun2023}
\begin{equation} \label{eq:rho}
    \rho_q(r) = \dfrac{\rho_{0,q}}{1+\exp\left(\dfrac{r-r_0}{a_0}\right)},
\end{equation}
where $q = n, p$ denotes neutrons and protons, respectively.
Two sets of parameters for the $^3\mathrm{He}$ projectile are considered as
\begin{itemize}
    \item Schechter and Canto (SC) density \cite{schechter1979}: $\rho_{0,n} = 0.0308$ fm$^{-3}$, $\rho_{0,p} = 0.0616$ fm$^{-3}$, $r_0 = 1.50$ fm, and $a_0 = 0.54$ fm, corresponding to a root-mean-square (rms) radius of 2.316~fm. 
    \item Ng\^{o} and Ng\^{o} (Ngo) density \cite{ngo1980}: $\rho_{0,n} = 0.0561$ fm$^{-3}$, $\rho_{0,p} = 0.1122$ fm$^{-3}$, $r_0 = 1.019$ fm, and $a_0 = 0.55$ fm, corresponding to an rms radius of 2.182~fm.
\end{itemize}
The experimental mass radius of $^3\mathrm{He}$ is reported to be $2.023$~fm in Ref.~\cite{morley2021}. The Ngo density provides better agreement with the experimental value.

The nucleon-nucleus potential $V_q(E,r)$ in Eq.~\eqref{eq:folding} is derived self-consistently from the Skyrme HF formalism after convergence of the mean-field calculation \cite{vautherin1972,colo2013}. It includes the central nuclear term, the spin-orbit interaction, and the Coulomb potential (only for protons)
\begin{equation} \label{eq:Vq}
    V_q(E,r) = V_0(E,r) + V_{\text{ls}}(r) + V_{\text{C}}(r).
\end{equation}

The energy-dependent central part of the potential in Eq.~\eqref{eq:Vq} is expressed as \cite{dover1971,dover1972}
\begin{align} \label{eq:V0}
V_0(E,r) &= \dfrac{m_q^*(r)}{m'} \bigg\{U_0(r) + \dfrac{1}{2} \dfrac{d^2}{dr^2}\left(\dfrac{\hbar^2}{2m_q^*(r)} \right) \nonumber \\
&- \dfrac{m_q^*(r)}{2\hbar^2} \left[\dfrac{d}{dr}\left(\dfrac{\hbar^2}{2m_q^*(r)}\right)\right]^2\bigg\} + \left[1-\dfrac{m^*_q(r)}{m'}\right]E_q, 
\end{align}
where $m' = mA/(A-1)$, with $m$ being the nucleon mass. The quantity $E_q$ represents the energy of an individual nucleon and, in this work, is approximated as $E_q \approx 0.467E$ in the c.m. frame.

The spin-orbit and Coulomb terms in Eq.~\eqref{eq:Vq} are given by
\begin{align} 
    V_{\text{ls}}(r) &= \dfrac{m_q^*(r)}{m'} U_{\text{ls}}(r)(\vec{\ell}\cdot\vec{s}), \label{eq:Vso} \\
    V_{\text{C}}(r) &= \dfrac{m_p^*(r)}{m'} U_{\text{C}}(r), \label{eq:VC}
\end{align}
where $U_0(r)$, $U_{\text{ls}}(r)$, $U_{\text{C}}(r)$, and the effective masses $m_q^*(r)$ are obtained from the self-consistent Skyrme HF equations \cite{vautherin1972}. Detailed procedures are provided in Refs.~\cite{dover1971,dover1972,anh2021CNO,anh2021C12O16,anh2022Li7Be7,anh2022Be10,anh2023O1415}. The charge, proton, and neutron radii of $^{4}$He obtained from the Skyrme HF calculation with the SLy4 interaction are $r_c = 2.122$~fm, $r_p = 1.967$~fm, and $r_n = 1.960$~fm, respectively.

In this work, scaling factors are introduced for different components of the potential. The factor $\lambda_0$ is applied to the central parts of the potential in Eq.~\eqref{eq:V0}, while $\lambda_{\text{ls}}$ is used to adjust the spin-orbit strength in Eq.~\eqref{eq:Vso} to reproduce the observed splitting of bound- and scattering-state partner levels. At large distances, the folded Coulomb potential in Eq.~\eqref{eq:VC} is smoothly matched to the asymptotic form $1/r$ to ensure the correct long-range behavior, with the matching radius chosen to be 10 fm. 

\section{Results and discussion} \label{sec:results}

\subsection{Low-energy $p+\alpha$  scattering}

The HF calculation is performed with the SLy4 interaction using the program described in Ref.~\cite{colo2013}. The HF potential in Eq.~\eqref{eq:Vq}, which corresponds to the real part of the optical potential, is employed to describe low-energy $p+\alpha$ scattering. Interestingly, the experimental phase shifts for both $s$- and $p$-wave states can be well reproduced by introducing a scaling factor $\lambda_0$ applied to the central component of the potential, without any modification to the spin-orbit term ($\lambda_{\text{ls}} = 1$).

\begin{figure}
    \centering
    \includegraphics[width=\linewidth]{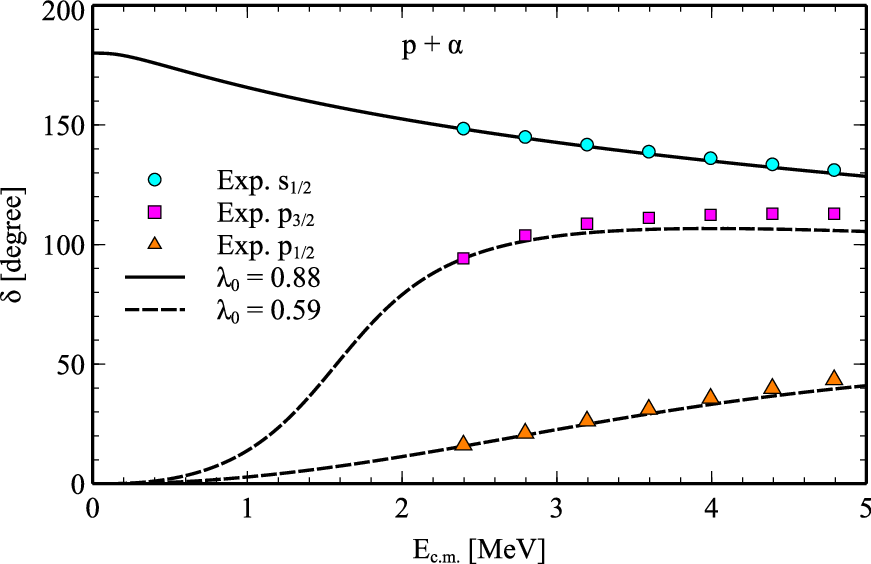}
    \caption{Calculated phase shifts for $p+\alpha$ elastic scattering below 5 MeV using the SLy4 interaction. The solid and dashed curves correspond to calculations with scaling factors $\lambda_{0}=0.88$ and $\lambda_{0}=0.59$ for the central potential, respectively. Experimental data for the $s_{1/2}$, $p_{3/2}$, and $p_{1/2}$ partial waves are taken from Ref.~\cite{schwandt1971}.}
    \label{fig:phase_palpha}
\end{figure}

Figure~\ref{fig:phase_palpha} shows the calculated phase shifts for $p+\alpha$ scattering at low energies for both $s$ and $p$ partial waves. The scaling factor $\lambda_0 = 0.88$ provides a good agreement with the experimental data of the $s_{1/2}$ phase shifts from Ref.~\cite{schwandt1971}. The scaling factor for $s$ scattering wave tends toward unity for heavier systems as discussed in previous studies~\cite{anh2021C12O16,anh2023O1415}, where the HF potential provides a more reliable description. Using $\lambda_0 = 0.88$ gives a proton radius of $r_p = 1.979$~fm, indicating that the scaling introduces only a minor change and does not significantly affect the nuclear structure of $^4\mathrm{He}$. In addition, applying $\lambda_0 = 0.59$ to the central potential well reproduces the $p_{1/2}$ and $p_{3/2}$ phase shifts without modifying the spin-orbit strength, indicating that the SLy4 spin-orbit interaction accurately describes the $p+\alpha$ system.

\subsection{Low-energy $^{3}\mathrm{He} + \alpha$ scattering}

\begin{figure}
    \centering
    \includegraphics[width=\linewidth]{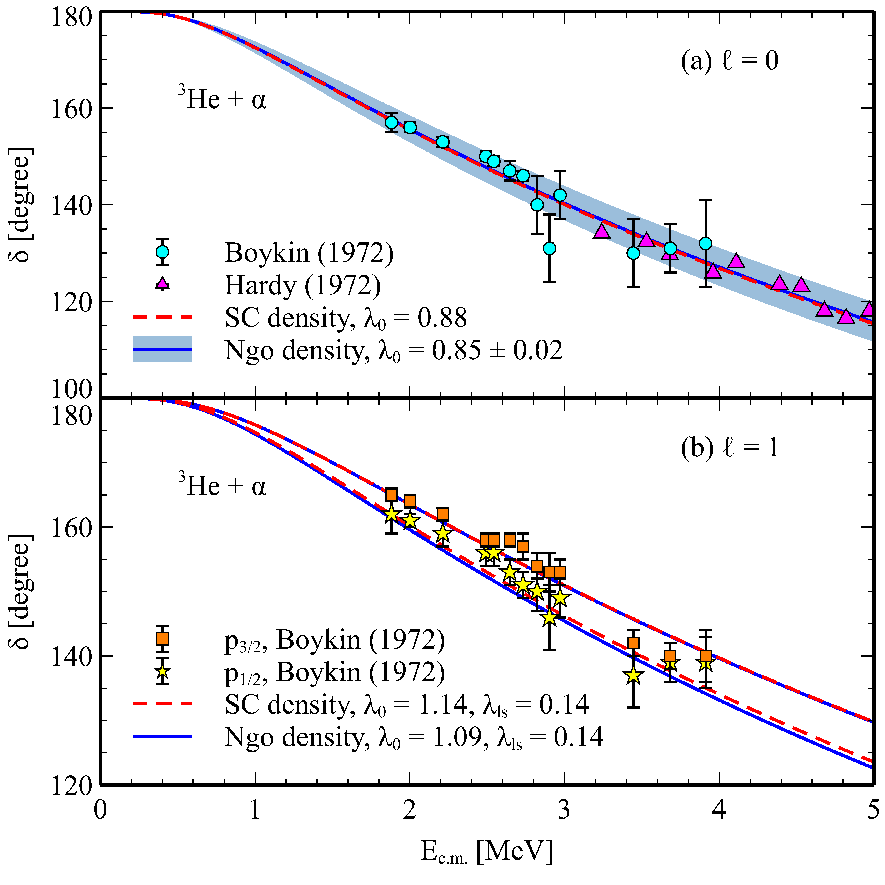}
    \caption{Calculated phase shifts for (a) $s$-wave ($\ell = 0$) and (b) $p$-wave ($\ell = 1$) $^{3}\mathrm{He}+\alpha$ elastic scattering below 5~MeV, obtained using folded SLy4 potentials constructed with different $^{3}\mathrm{He}$ density distributions. The shaded band in (a) corresponds to the uncertainty range $\Delta\lambda_{0} = \pm 0.02$ around the central scaling factors $\lambda_{0}$. Experimental data are taken from Refs.~\cite{boykin1972,hardy1972}.}
    \label{fig:phase_3He_alpha}
\end{figure}

Figure~\ref{fig:phase_3He_alpha}(a) shows the calculated $s$-wave phase shifts for $^{3}\mathrm{He}+\alpha$ elastic scattering below 5~MeV, obtained using the folded potentials constructed with the scaling factors $\lambda_0 = 0.88$ using SC density and $\lambda_0 = 0.85$ using the Ngo density. The resulting phase shifts are shown to be the same for both densities. These exact values of $\lambda_0$ are chosen to reproduce the lowest-energy experimental data points, and the corresponding uncertainty of $\Delta\lambda_0$ is estimated to be $\pm 0.02$ for both cases. The shaded band represents the uncertainties in the calculations when reproducing the experimental data from Refs.~\cite{boykin1972,hardy1972}. The calculated results agree well with the experimental phase shifts, particularly at low energies. The optimal scaling factors $\lambda_0$ extracted from the $^{3}\mathrm{He}+\alpha$ analysis are found to be close to those determined from $p+\alpha$ scattering. This result demonstrates that the $^{3}\mathrm{He}+\alpha$ elastic scattering can be reliably described once an accurate $p+\alpha$ optical potential and a realistic $^{3}\mathrm{He}$ density are used.

\begin{table}[b]
\setlength{\tabcolsep}{10pt}
\centering
\caption{Scaling factors $\lambda_0$ and $\lambda_{\text{ls}}$ used for the scattering and bound states of the $^{3}\mathrm{He}+\alpha$ system. Results are shown for both the SC and Ngo density distributions employed in the folding procedure.}
\label{tab:lambda}
\begin{tabular}{lcccc}
\hline\hline
 & \multicolumn{2}{c}{SC density} & \multicolumn{2}{c}{Ngo density} \\ \hline
 Scattering states & $\lambda_0$ & $\lambda_{\text{ls}}$ & $\lambda_0$ & $\lambda_{\text{ls}}$ \\ \hline
 $s_{1/2}$, $d_{5/2}$, $d_{3/2}$ & 0.88 & 0.14 & 0.85 & 0.14 \\
 $p_{3/2}$, $p_{1/2}$ & 1.14 & 0.14 & 1.09 & 0.14 \\ \hline
 Bound states & $\lambda_0$ & $\lambda_{\text{ls}}$ & $\lambda_0$ & $\lambda_{\text{ls}}$ \\ \hline
 $p_{3/2}$, $p_{1/2}$ & 1.07 & 0.09 & 1.04 & 0.08 \\ \hline\hline
\end{tabular}
\end{table}

Figure~\ref{fig:phase_3He_alpha}(b) shows the calculated phase shifts for the scattering $p$ states. The spin-orbit scaling factor $\lambda_{\text{ls}} = 0.14$, used for both the SC and Ngo densities, is primarily constrained by the experimental phase shifts of the $p_{1/2}$ and $p_{3/2}$ states reported in Refs.~\cite{boykin1972,hardy1972}. The value $\lambda_{\text{ls}}=0.14$ is also applied to the $d_{5/2}$ and $d_{3/2}$ partial waves. The available $d$-wave phase-shift data exhibit larger uncertainties and do not show a clear spin-orbit splitting~\cite{hardy1972,spiger1967}. The strength of the folded spin-orbit potential in the $^{3}\mathrm{He}+\alpha$ system is found to be roughly an order of magnitude weaker than that in the $p+\alpha$ case. The same central scaling factors $\lambda_0$ are applied to the $s$ and $d$ states, consistent with Refs.~\cite{mohr1993,mohr2009}. This assumption is further supported by the nearly same phenomenological Gaussian or Woods-Saxon potential depths for the $s$ and $d$ scattering states without spin-orbit terms reported in Refs.~\cite{xu2013,tursunov2018}. The $\lambda_0$ values used for the partial waves are summarized in Table~\ref{tab:lambda}.

\begin{figure}
    \centering
    \includegraphics[width=\linewidth]{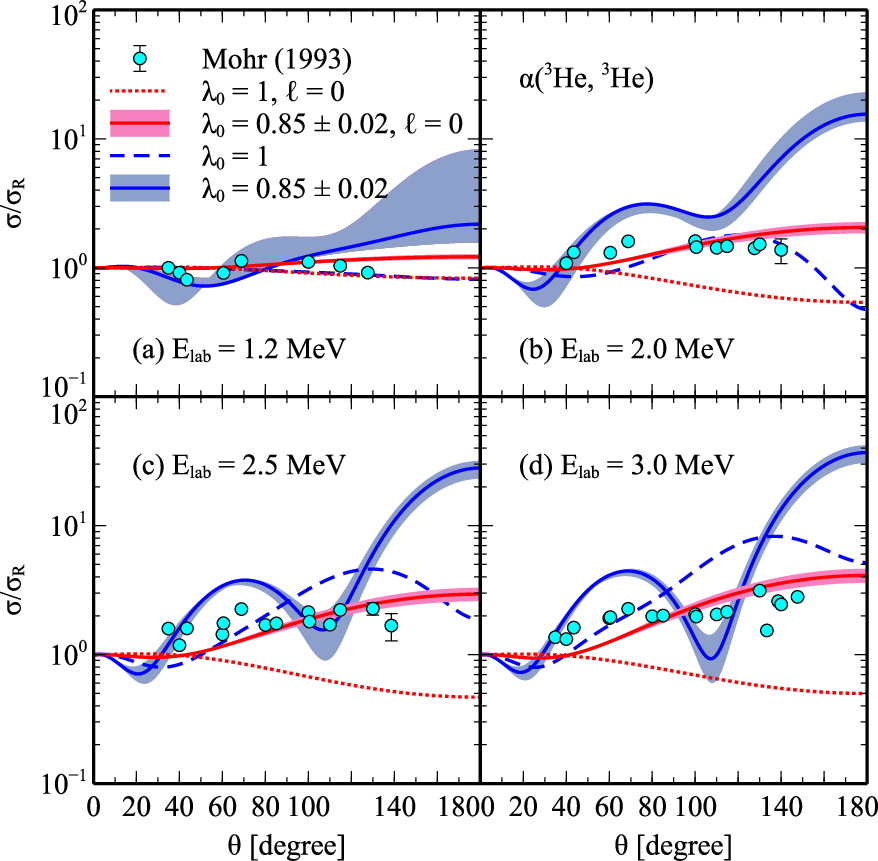}
    \caption{Calculated angular distributions of $^{3}\mathrm{He}+\alpha$ elastic scattering at $E_{\text{lab}}(^{3}\mathrm{He})=$ (a) 1.2 MeV, (b) 2.0 MeV, (c) 2.5 MeV, and (d) 3.0 MeV, normalized to the Rutherford cross section. Results including all partial waves (blue) and only the $\ell=0$ component (red) are shown. Solid curves (with shaded bands) are calculated using scaled folded potentials with $\lambda_{0}=0.85\pm 0.02$ and $\lambda_{\text{ls}} =0.14$, while dashed curves correspond to unscaled results ($\lambda_{0}=\lambda_{\text{ls}}=1$). All calculations use the Ngo density. Experimental data are from Ref.~\cite{mohr1993}.}
    \label{fig:cs_3He_alpha}
\end{figure}

Figures~\ref{fig:cs_3He_alpha}(a-d) show the $^{3}\mathrm{He}+\alpha$ elastic-scattering angular distributions normalized to the Rutherford cross section for energies between $1.2$ MeV and $3.0$~MeV. The calculations include either all partial waves (blue) or only the $\ell=0$ component (red) in the partial-wave expansion. The scaled potentials with $\lambda_0=0.85$ and $\lambda_{\text{ls}}=0.14$ provide a clear improvement over the unscaled case ($\lambda_0=\lambda_{\text{ls}}=1$), particularly at large scattering angles. No fine-tuning of $\lambda_0$ was performed; the goal is to demonstrate that $\lambda_0 = 0.85 \pm 0.02$ already reproduces the main experimental 
trends when the dominant $\ell=0$ contribution is considered. At very low energies, the cross section is dominated by the  $\ell=0$ contribution, which is directly constrained by the $s$-wave phase shifts. Further refinement could be achieved using separate potentials for each partial wave as discussed in Ref.~\cite{mohr1993}, though this is beyond the scope of the present work.

\subsection{Astrophysical $S$ factor of $^{3}\mathrm{He}(\alpha,\gamma)^{7}\mathrm{Be}$ reaction}

The bound states of $^{7}$Be are obtained by solving the cluster relative-motion Schr\"odinger equation using the folded $^{3}\mathrm{He}+\alpha$ potential. The nucleus $^{7}\mathrm{Be}$ can be described within a two-cluster model as a system of $^{3}\mathrm{He}$ (with spin $J_a^\pi = 1/2^+$) and $\alpha$ ($J_A^\pi = 0^+$). Below the $^{3}\mathrm{He}+\alpha$ threshold ($Q_\alpha = 1.587$~MeV), two bound states are the ground state with spin-parity $J^\pi_B = 3/2^-$ and the first excited state with $J^\pi_B = 1/2^-$ at an excitation energy of $E_x = 429$~keV. In this description, the ground and first excited states correspond to $^{3}\mathrm{He}$ occupying the $2p_{3/2}$ and $2p_{1/2}$ cluster states, respectively~\cite{mohr2009}. The allowed cluster states are constrained by the Wildermuth condition $Q=2N+L$. For $^{3}$He, $Q = \sum_{i=1}^{3}(2n_i+\ell_i) = 3$, giving $(N,L) = (1,1)$ as the lowest configuration. With $L = 1$ and the intrinsic spin $1/2$ of the $A=3$ cluster, spin-orbit coupling produces the doublet $J = L \pm 1/2$, corresponding to $3/2^-$ and $1/2^-$.

The binding energies of the $2p_{3/2}$ and $2p_{1/2}$ states, given by $E_B = -Q_\alpha + E_x$, are $-1.587$~MeV and $-1.158$~MeV, respectively. To reproduce these binding energies, the scaling factors $\lambda_0$ and $\lambda_{\text{ls}}$ for the central and spin-orbit components of the folded potential were adjusted simultaneously. The best agreement was obtained with $\lambda_0 = 1.07$ and $\lambda_{\text{ls}} = 0.09$ when using the SC density, and with $\lambda_0 = 1.04$ and $\lambda_{\text{ls}} = 0.08$ when using the Ngo density for the bound-state $^{3}\mathrm{He}+\alpha$ potential as shown in Table~\ref{tab:lambda}. It is worth noting that the adopted spin-orbit scaling factors are nearly identical to the spin-orbit normalization parameters reported in Ref.~\cite{mohr1993}.

\begin{table}[b]
    \centering
    \setlength{\tabcolsep}{3pt}
    \caption{Asymptotic normalization coefficients (ANCs) in fm$^{-1/2}$ for the bound $p_{3/2}$ and $p_{1/2}$ cluster states in $^{7}$Be, extracted from the folded Skyrme potentials using different $^{3}$He densities. Empirical values obtained from indirect methods are included for comparison.}
    \label{tab:ANCs}
    \begin{tabular}{lll} \hline\hline 
        Source & $C_{p_{3/2}}$ & $C_{p_{1/2}}$ \\ 
         & [fm$^{-1/2}$] & [fm$^{-1/2}$] \\ \hline
        Tursunmahatov \textit{et al.} (2012)~\cite{tursunmahatov2012} & $4.827^{+0.102}_{-0.244}$ & $3.987^{+0.075}_{-0.193}$ \\
        Yarmukhamedov \textit{et al.} (2016)~\cite{yarmukhamedov2016}            & $4.785^{+0.073}_{-0.073}$   & $4.243^{+0.059}_{-0.059}$   \\
        Kiss \textit{et al.} (2020)~\cite{kiss2020} & $4.565^{+0.121}_{-0.124}$ & $3.586^{+0.069}_{-0.070}$ \\ \hline 
        Present work (SC density)      & $5.018$   & $4.256$   \\
        Present work (Ngo density)     & $4.766$   & $4.056$   \\ 
        \hline \hline
    \end{tabular}
\end{table}

The asymptotic normalization coefficients (ANCs, in fm$^{-1/2}$) are also extracted from the present model. Table~\ref{tab:ANCs} shows the ANC values obtained from the folded Skyrme potentials using both density distributions. The results obtained with the Ngo density show better agreement with the experimental data \cite{tursunmahatov2012,yarmukhamedov2016}, differing by less than 3\% compared to Ref.~\cite{tursunmahatov2012}, although they remain slightly higher than the values from Ref.~\cite{kiss2020}. Within the potential-model framework, the ANCs provide a stringent constraint on the interaction parameters, as also emphasized in Ref.~\cite{tursunov2018}.

\begin{figure}[]
    \centering
    \includegraphics[width=\linewidth]{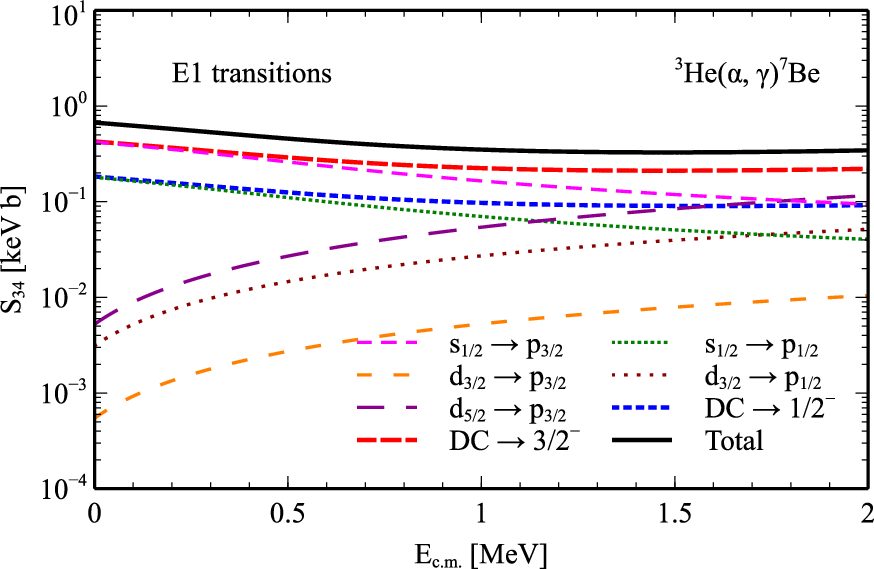}
    \caption{Calculated $E1$ transitions for direct capture (DC) to the ground state ($3/2^-$) and first excited state ($1/2^-$) of $^{7}\mathrm{Be}$ using the Ngo density in the folded potential. At low energies, the capture is dominated by transitions from $s$-wave scattering states.}
    \label{fig:E1transitions}
\end{figure}

This work focuses exclusively on the $E1$ transition, which provides the dominant contribution to the radiative-capture cross section. Higher-order transitions such as $E2$ and $M1$, as discussed in Refs.~\cite{mohr1993,tursunov2018}, are negligible at the low energies of astrophysical interest. The $E1$ transition originates primarily from the captured $s$ and $d$ scattering waves. As shown in Table~\ref{tab:lambda}, the central scaling factors $\lambda_0 = 0.88$ (SC) and $\lambda_0 = 0.85$ (Ngo) are applied to both the $s$- and $d$-wave potentials. The spin-orbit component of the folded potential is fixed at $\lambda_{\text{ls}} = 0.14$.

Figure~\ref{fig:E1transitions} shows the calculated possible $E1$ transitions from scattering $s$ and $d$ states to the bound $p$ states, obtained using the folded potential constructed with the Ngo density. The transition to the ground state ($3/2^-$) dominates over that to the first excited state ($1/2^-$). The calculated branching ratio, defined as $R = \sigma_{429}/\sigma_0$, with $\sigma_0$ 
and $\sigma_{429}$ denoting the capture cross sections to the ground and first excited states, respectively, is $0.43$ for both densities at very low energies, indicating that it is essentially insensitive to the choice of projectile density. This observation is consistent with the conclusion of Ref.~\cite{mohr2009}, which reported that $R$ is largely unaffected by variations in the potential strength. At energies below 1~MeV, the capture process at low energies is dominated by  $s \to p$ transitions, while above 1~MeV the $d \to p$ contributions become increasingly significant and eventually comparable in strength. 

\begin{figure}[]
    \centering
    \includegraphics[width=\linewidth]{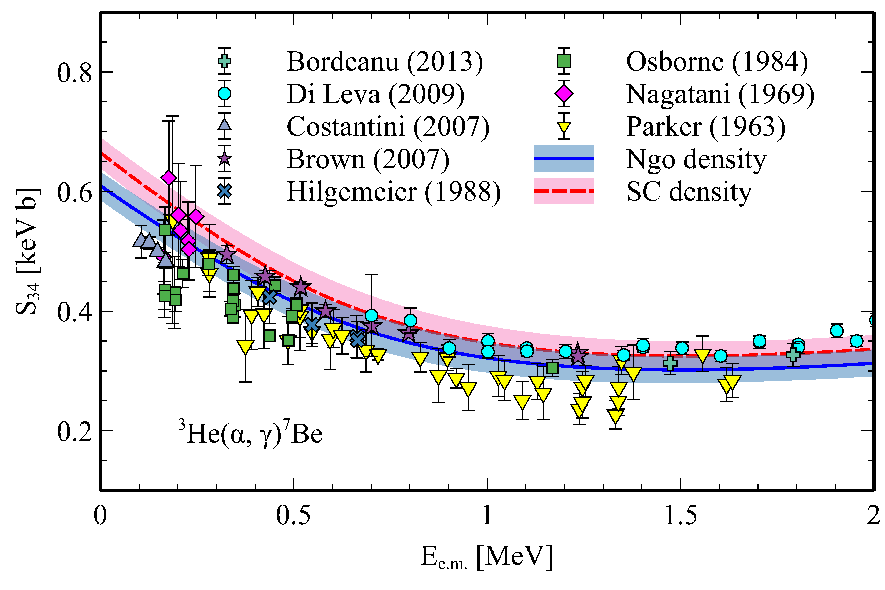}
    \caption{Calculated astrophysical $S_{34}$ factor for the $^{3}\mathrm{He}(\alpha,\gamma)^{7}\mathrm{Be}$ reaction below 2 MeV using folded potentials constructed the SC and Ngo $^{3}\mathrm{He}$ densities. The shaded bands represent the uncertainties. The results are compared with experimental data from Refs.~\cite{bordeanu2013, dileva2009, costantini2007, costantini2008, brown2007, hilgemeier1988, osborne1984, nagatani1969, parker1963}. The calculation based on the Ngo density provides better overall agreement with the experimental data.}
    \label{fig:sfactor}
\end{figure}

Figure~\ref{fig:sfactor} presents the calculated low-energy astrophysical $S$ factor for the $^{3}\mathrm{He}(\alpha,\gamma)^{7}\mathrm{Be}$ reaction using different $^{3}\mathrm{He}$ densities in the folded potentials, compared with experimental data from Refs.~\cite{ bordeanu2013, nagatani1969, dileva2009, costantini2007, costantini2008, brown2007, hilgemeier1988, osborne1984, parker1963}. The uncertainty in the astrophysical $S$ factor comes mainly from the scaling factor $\lambda_0$ used to reproduce the $s_{1/2}$ phase shifts as shown in Fig.~\ref{fig:phase_3He_alpha}, with $\Delta\lambda_0 = \pm 0.02$ in both cases. The two calculations employing the SC and Ngo densities yield slightly different results. Overall, the prediction obtained with the Ngo density shows better agreement with most experimental datasets. This improvement can be attributed to the smaller rms radius associated with the Ngo density, which provides a more realistic description of the spatial distribution of the $^{3}\mathrm{He}$ cluster and results in ANCs closer to experimental values. In contrast, the $S$ factor calculated with the SC density is about 8\% higher but remains consistent with the experimental data reported in Refs.~\cite{brown2007,nagatani1969,dileva2009}. At low energies, the imaginary part of the optical potential is negligible, while at higher energies, additional reaction channels and coupling effects become important \cite{hao2015}. The present approach, therefore, provides a reliable description of the reaction in the astrophysically relevant energy range.

\begin{table}[]
    \centering
    \setlength{\tabcolsep}{5pt}
    \caption{Comparison of available $S_{34}(0)$ values in keV~b for the $^3\mathrm{He}(\alpha,\gamma)^7\mathrm{Be}$ reaction at threshold.}
    \label{tab:S0}
    \begin{tabular}{ll} \hline \hline 
        References & $S_{34}(0)$ \\ 
         & [keV~b]  \\ \hline
        Nagatani \textit{et al.} (1969) \cite{nagatani1969} & $0.58 \pm 0.07$ \\
        Robertson \textit{et al.} (1983) \cite{robertson1983} & $0.63 \pm 0.04$ \\
        Brown \textit{et al.} (2007) \cite{brown2007} & $0.595 \pm 0.018$ \\
        Cyburt \& Davids (2008) \cite{cyburt2008} & $0.580 \pm 0.043$ \\ 
        Costantini \textit{et al.} (2008) \cite{costantini2008} & $0.567 \pm 0.018$ \\
        Mohr (2009) \cite{mohr2009} & $0.53 \pm 0.05$ \\
        Di Leva \textit{et al.} (2009) \cite{dileva2009} & $0.590 \pm 0.016$ \\
        Neff (2011) \cite{neff2011} & $0.593$ \\
        Dohet-Eraly \textit{et al.} (2016) \cite{dohet2016} & $0.59$ \\
        Tak{\'a}cs \textit{et al.} (2018) \cite{takacs2018} & $0.598 \pm 0.051$ \\
        Zhang \textit{et al.} (2020) \cite{zhang2020} & $0.577^{+0.015}_{-0.016}$ \\
        Acharya \textit{et al.} (2025) \cite{acharya2025} & $0.561 \pm 0.028$ \\
        Atkinson \textit{et al.} (2025) \cite{atkinson2025} & $0.545(1)$ \\
        \hline
        Present work (SC density) & $0.666 \pm 0.026$ \\
        Present work (Ngo density, \textit{recommended}) & $0.610 \pm 0.024$ \\\hline\hline
    \end{tabular}
\end{table}

The extrapolated values are $S_{34}(0) = 0.666 \pm 0.026$~keV~b (SC) and $0.610 \pm 0.024$~keV~b (Ngo), the latter in excellent agreement with previous evaluations and adopted as the recommended result. It also lies near the upper limit of the most recent evaluation reported Solar Fusion III~\cite{acharya2025}. The present results are consistent with state-of-the-art theoretical predictions, including the \textit{ab initio} no-core shell model with continuum~\cite{dohet2016}, the fermionic molecular dynamics model~\cite{neff2011}, and the halo effective
field theory \cite{zhang2020}. It is worth emphasizing that this approach does not employ any spectroscopic factors, in contrast to the potential-model approaches used in Refs.~\cite{mohr1993,mohr2009,xu2013}.

A quadratic polynomial fits the calculated $S$ factors up to 500~keV. The slope parameter is $S'_{34}(0)/S_{34}(0) = -0.754 \pm 0.051$~MeV$^{-1}$, which is in agreement with the value of $-0.92\pm 0.18$ MeV$^{-1}$ reported in Ref.~\cite{cyburt2008}. This result is also consistent with the microscopic potential-model calculation using a double-folding approach presented in Ref.~\cite{mohr2009}, which obtained $S'_{34}(0)/S_{34}(0) = -0.73$~MeV$^{-1}$. At the Gamow peak energy, the experimental value $S_{34}(23^{+6}_{-5}~\text{keV}) = 0.548 \pm 0.054$~keV~b inferred from solar neutrino fluxes~\cite{takacs2015} is well reproduced by our calculation, which yields $S_{34}(23\pm 6~\text{keV}) = 0.600 \pm 0.027$~keV~b. To compare with other references, we listed the selected $S_{34}(0)$ values in Table~\ref {tab:S0}. No fitting procedure was applied to optimize the scaling factor $\lambda_0$ for the astrophysical $S$ factor; instead, the same value was retained throughout to ensure internal consistency, maintain independence from experimental adjustments, and demonstrate the predictive power of the present approach. 

\section{Conclusion\label{sec:conclusion}}
A unified microscopic description of light-ion elastic scattering and radiative-capture reactions has been achieved within the Skyrme HF framework. Using a self-consistent potential derived from the SLy4 interaction, both the $p+\alpha$ and $^{3}\mathrm{He}+\alpha$ systems are simultaneously described with a small number of scaling adjustments. The calculated phase shifts, cross sections, and astrophysical $S$ factor show good agreement with experimental data. The extrapolated $S_{34}(0)$ value is found to be sensitive to the choice of the $^{3}\mathrm{He}$ density employed in the folding procedure. Among the tested densities, the Ngo parametrization provides the most consistent results, yielding rms radii and ANCs for the bound states closest to experiment. The spin-orbit strength in the two-cluster $^{3}\mathrm{He}+\alpha$ system is found to be significantly reduced compared to the nucleon-nucleus case.

Overall, the present work demonstrates that the Skyrme HF framework provides a consistent microscopic approach for describing both nucleon-nucleus and nucleus-nucleus interactions. The present framework will be applied to the $^{7}\mathrm{Li}+\alpha$ channel~\cite{tang2025}, which lies 2.56~MeV below the proton-emission threshold of $^{11}$B, where an $s$-state resonance has been identified near the threshold within the Skyrme HF formalism~\cite{anh2022Be10}. The approach can also be extended to heavy-ion systems of astrophysical relevance, providing reliable predictive power with minimal empirical input.

\section*{Acknowledgements}
The authors would like to thank Bui Minh Loc (SDSU) for valuable discussions. This research is funded by the Vietnam National Foundation for Science and Technology Development (NAFOSTED) under grant number 103.04-2025.07.

\bibliographystyle{apsrev4-2}
\bibliography{refs.bib}

\end{document}